\documentclass[runningheads]{llncs}

\usepackage[pagebackref=false,breaklinks=true,colorlinks=true,bookmarks=false,citecolor=blue]{hyperref}
\usepackage{amsmath}
\usepackage{amssymb}
\usepackage{graphicx}
\usepackage{url}
\newcommand{\eg}[1]{\textit{e.g.,}}
\newcommand{\ie}[1]{\textit{i.e.,}}
\newcommand{\etc}[1]{\textit{etc}}
\newcommand{\figref}[1]{Fig.\!~\ref{#1}}

\usepackage{threeparttable}
\usepackage{color}
\usepackage{multirow}
\DeclareMathOperator*{\Motimes}{\text{\raisebox{0.25ex}{\scalebox{0.6}{$\bigotimes$}}}}

\DeclareMathOperator*{\Moplus}{\text{\raisebox{0.25ex}{\scalebox{0.6}{$\bigoplus$}}}}
\usepackage[pagebackref=false,breaklinks=true,colorlinks=true,bookmarks=false,citecolor=blue]{hyperref}
\begin{document}
\title{Task Transformer Network for Joint MRI Reconstruction and Super-Resolution}  
%
%
\author{Chun-Mei Feng\inst{1,2}
\and
Yunlu Yan\inst{1}
\and
Huazhu Fu\inst{2}
\and
Li Chen\inst{3}
\and
Yong Xu \inst{1}}
\institute{$^1$ Shenzhen Key Laboratory of Visual Object Detection and Recognition, Harbin Institute of Technology, shenzhen\\
$^2$ Inception Institute of Artificial Intelligence \\
$^3$ First Affiliated Hospital with Nanjing Medical University \\
{\tt \{strawberry.feng0304\}@gmail.com}\\
{\tt \href{https://github.com/chunmeifeng/T2Net}{https://github.com/chunmeifeng/T2Net}}
\thanks{C.-M.~Feng and Y.~Yan are contributed equally to this work. This work was done during the internship of C.-M.~Feng at Inception Institute of Artificial Intelligence. Yong Xu is the corresponding author.}
}

%
\maketitle              
\begin{abstract}
%
The core problem of Magnetic Resonance Imaging (MRI) is the trade off between acceleration and image quality. Image reconstruction and super-resolution are two crucial techniques in Magnetic Resonance Imaging (MRI). Current methods are designed to perform these tasks separately, ignoring the correlations between them. In this work, we propose an end-to-end task transformer network (T$^2$Net) for joint MRI reconstruction and super-resolution, which allows representations and feature transmission to be shared between multiple task to achieve higher-quality, super-resolved and motion-artifacts-free images from highly undersampled and degenerated MRI data. Our framework combines both reconstruction and super-resolution, divided into two sub-branches, whose features are expressed as queries and keys. Specifically, we encourage joint feature learning between the two tasks, thereby transferring accurate task information. We first use two separate CNN branches to extract task-specific features. Then, a task transformer module is designed to embed and synthesize the relevance between the two tasks. Experimental results show that our multi-task model significantly outperforms advanced sequential methods, both quantitatively and qualitatively.


\end{abstract}
\section{Introduction}
Magnetic resonance imaging (MRI) is a popular diagnostic modality. However, the physics behind its data acquisition process makes it inherently slower than other methods such as computed tomography\!~(CT) or X-rays. Therefore, improving the acquisition speed of MRI has been an important research goal for decades. MRI reconstruction and super-resolution (SR) are two main methods for this, where the former accelerates MRI by reducing the $k$-space sampling rate, and the latter achieves a high-resolution\!~(HR) image by restoring a single degenerated low-resolution\!~(LR) image~\cite{feng2021brain}.

Outstanding contributions have been made in both areas~\cite{feng2021MINet,feng2021DONet,feng2021accelerated}. Specifically, compressed sensing (CS)~\cite{lai2016image}, low-rank~\cite{shin2014calibrationless}, dictionary learning~\cite{ravishankar2010mr,zhan2015fast}, 
and manifold fitting~\cite{nakarmi2017kernel} techniques utilize various priors to overcome aliasing artifacts caused by the violation of the Shannon-Nyquist sampling theorem for MRI reconstruction. With the renaissance of deep neural networks, different convolutional neural network (CNN) approaches have also been developed for fast MRI reconstruction~\cite{zhu2018image,feng2021DualOctConv}. Typical examples include model-based unrolling methods, \eg, VN-Net~\cite{hammernik2018learning}, which generalizes the CS formulation to a variational model, and ADMM-Net, which is derived from the iterative procedures~\cite{yang2016deep}; end-to-end learning methods, \eg, using U-Net as the basic framework to solve the problem of MRI reconstruction~\cite{qin2018convolutional,huang2019mri}; and generative adversarial networks (GANs)~\cite{yang2017dagan,mardani2018deep}. In addition to various network structures, series of convolutions based on the characteristics of MRI data have also been designed to solve the problem of MRI reconstruction~\cite{feng2021DualOctConv,wang2020deepcomplexmri}. For MRI SR, iterative algorithms (\eg, low rank or sparse representation) take image priors into account  as regularization items and try to obtain a higher-quality image from a single LR image~\cite{wang2014sparse,zhang2007application}.
Similarly, CNN approaches have achieved state-of-the-art
performance in SR~\cite{chaudhari2018super,lyu2019super}. For example, residual learning can be used to extract multi-scale information and obtain higher-quality images~\cite{shi2018super,oktay2016multi}. GAN-based methods have also been used to recover HR details from an LR input~\cite{chen2018efficient,mahapatra2019image}. 



However, these works are designed to perform one specific function, \ie, train a single model to carry out the desired task. While acceptable performance can be achieved in this way, information that might help the model perform better in certain metrics is often ignored, since too much focus is given to one single task. In the real world, a network that can perform multiple tasks simultaneously is far preferable to a set of independent networks, as it can provide a more complete visual system. Since related tasks often share features, real-world tasks tend to have strong dependencies. Recently, multi-task learning has been successfully applied to various fields, including natural language processing~\cite{collobert2008unified}, speech recognition~\cite{kim2017joint} and computer vision~\cite{liu2019end}. By sharing representations between related tasks, the model can better generalize to the original task. Compared with standard single-task learning, multi-task models can express both shared and task-specific characteristics. In natural images, multi-task learning has been widely used for image enhancement~\cite{cai2019fcsr,zhang2018joint}. However, current models directly incorporate different tasks into the network in a sequential manner, without exploring the features shared across the tasks.
Inspired by the powerful visual capabilities of transformer and multi-task learning, we propose an end-to-end task transformer network, named T$^2$Net, for multi-task learning, which integrates both MRI reconstruction and SR. Our contributions are three-fold: \textbf{First}, to the best of our knowledge, we are the first to introduce the transformer framework into multi-task learning for MRI reconstruction and SR. Our network allows representations to be shared between the two tasks, leveraging knowledge from one task to speed up the learning process in the other and increase the flexibility for sharing complementary features. \textbf{Second}, we develop a framework with two branches for expressing task-specific features and a task transformer module for transferring shared features. More specifically, the task transformer module includes relevance embedding, transfer attention and soft attention, which enable related tasks to share visual features. \textbf{Third}, we demonstrate that our multi-task model generates superior results compared to various sequential combinations of state-of-the-art MRI reconstruction and super-resolution models. 

\section{Method}
\subsection{Task Transformer Network}
Let $\mathbf{y}$ be the complex-valued, fully sampled $k$-space. The corresponding fully sampled HR image with a size of $h \times w$ can be obtained by $\mathbf{x}=\mathcal{F}^{-1}(\mathbf{y})$, where $\mathcal{F}^{-1}$ is the inverse 2D fast Fourier transform (FFT). To accelerate the MRI acquisition, a binary mask operator $M$ defining the Cartesian acquisition trajectory is used to select a subset of the $k$-space points. Therefore, the undersampled $k$-space is obtained by $\hat{\mathbf{y}}= M \odot \mathbf{y}$, where $\odot$ denotes element-wise multiplication. Accordingly, the zero-filled image can be  expressed as $\hat{\mathbf{x}}=\mathcal{F}^{-1}(\hat{\mathbf{y}})$. In MRI super-resolution, to obtain the LR image $\mathbf{x}_{LR}$ with a size of $\frac{h}{s}\!\times\!\frac{w}{s}$ ($s$ is the scale factor), we follow~\cite{chen2018efficient}, first downgrading the resolution by truncating the outer part of $\mathbf{y}$ with a desired factor to obtain $\mathbf{y}_{LR}$, and then applying $\mathcal{F}$ to it. Therefore, if we apply downgrading to $\hat{\mathbf{x}}$, we will obtain the undersampled, degenerated MRI data for our multi-task input $\hat{\mathbf{x}}_{LR}$.




To effectively achieve higher-quality, motion-artifact-free images from highly undersampled and degenerated MRI data $\hat{ \mathbf{x}}_{LR}$, we propose a simple and effective end-to-end framework, named the Task Transformer Network (T$^2$Net). As shown in \figref{fig1}, our multi-task framework consists of three parts: an SR branch, a reconstruction (Rec) branch and a task transformer module. The first two branches are used to extract task-specific features, providing our network the ability to learn features tailored to each task. The task transformer module is then used to learn shared features, encouraging the network to learn a generalizable representation. As can be seen, the input of the two branches is the undersampled, degenerated MRI data $\hat{\mathbf{x}}_{LR}$, which contains motion artifacts and blurring effects. The output of the Rec branch is the LR motion-free image ${\mathbf{x}'}_{LR}$, while the output of the SR branch is our final desired high-quality, super-resolved and motion-free image $\mathbf{x'}$. Our framework can be approximated using neural networks by minimizing an $\ell_1$ loss function:
\begin{equation}
\small{
\hat{\theta}=\underset{\theta_{1},\theta_{2}}{\arg \min } \sum_{j}^{N}\left(\alpha\left\|\mathbf{x}^{j}-f_{c n n}^{SR}\left(\hat{ \mathbf{x}}^{j}_{LR} \mid \theta_{1}\right)\right\|_{1}+\beta\left\|\mathbf{x}_{LR}^{j}-f_{c n n}^{Rec}\left(\hat{ \mathbf{x}}^{j}_{LR} \mid \theta_{2}\right)\right\|_{1}\right),
}
\end{equation}
where $f_{c n n}^{SR}$ and $f_{c n n}^{Rec}$ represent the mapping functions of the SR and Rec branches with parameters $\theta_{1}$ and $\theta_{2}$, respectively, and $\alpha$ and $\beta$ are used to balance the weights of the two branches. Note that with sufficient training data
$\{\mathbf{x}^{j}, \hat{ \mathbf{x}}^{j}_{LR}\}$
and the SGD algorithm, we can obtain well-trained weights $\hat{\theta}$.

\subsubsection{SR Branch.}
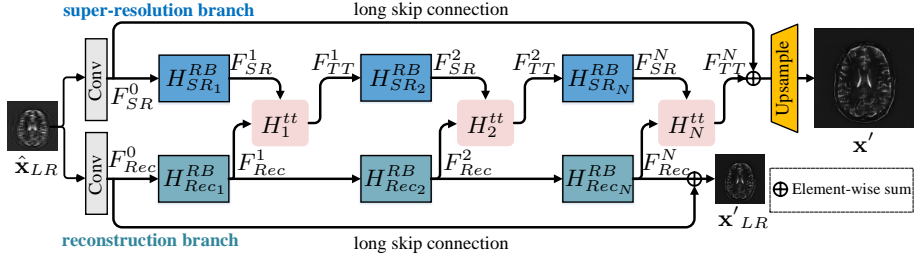
\begin{figure}[!t]
\centering
  \includegraphics[width=0.99\textwidth]{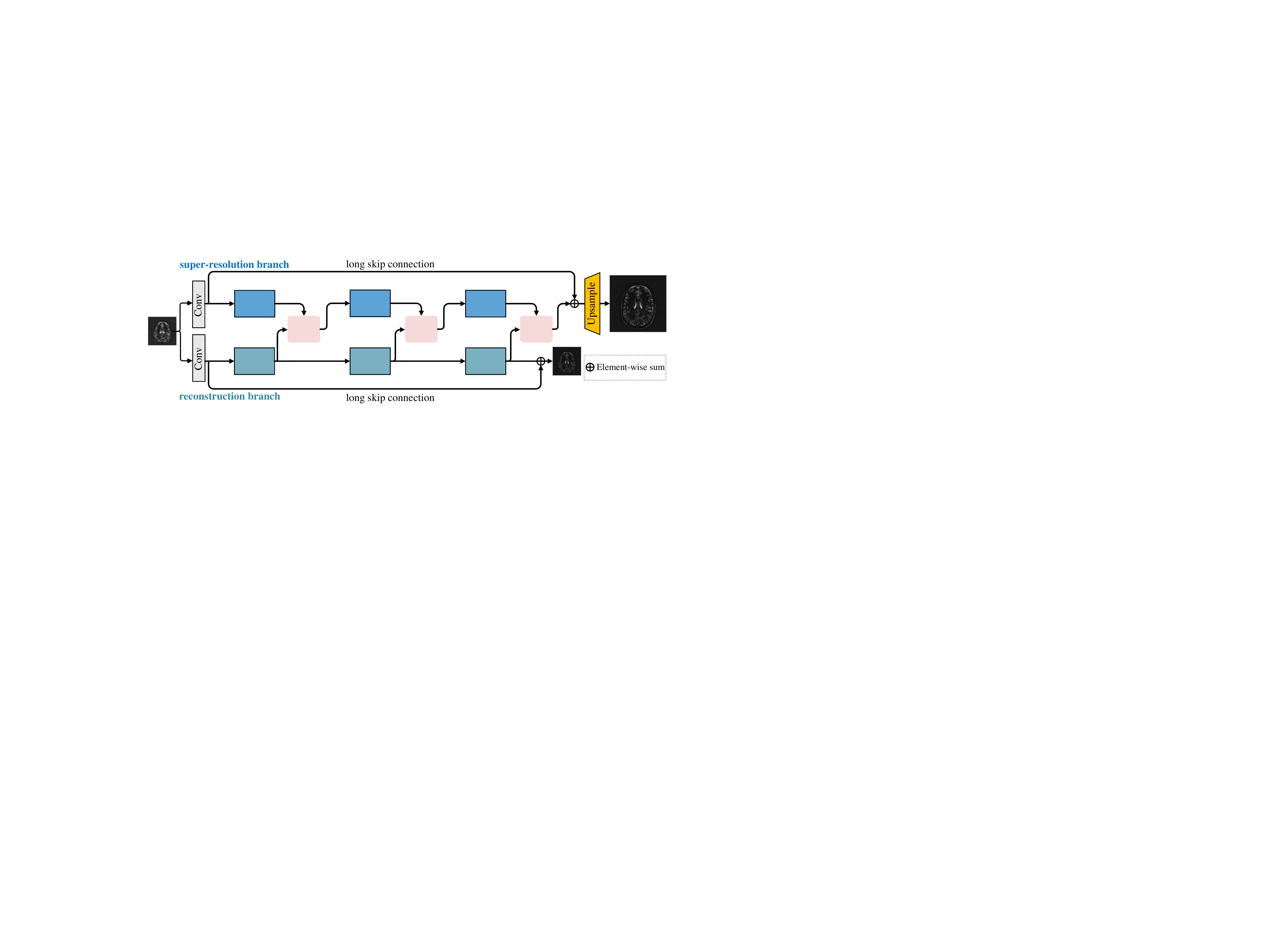}
  \put(-340,31){\footnotesize $\hat{ \mathbf{x}}_{LR}$}
  \put(-74,11){\footnotesize ${ \mathbf{x}'}_{LR}$}  
  \put(-23,39){\footnotesize ${ \mathbf{x}}'$}  
  \put(-283,64){\footnotesize $H^{RB}_{SR_{1}}$} 
  \put(-206,64){\footnotesize $H^{RB}_{SR_{2}}$} 
  \put(-131,64){\footnotesize $H^{RB}_{SR_{N}}$} 
  \put(-284,26){\footnotesize $H^{RB}_{Rec_{1}}$}
  \put(-208,26){\footnotesize $H^{RB}_{Rec_{2}}$} 
  \put(-132,26){\footnotesize $H^{RB}_{Rec_{N}}$} 
  \put(-247,46){\footnotesize $H_{1}^{tt}$} 
  \put(-169,46){\footnotesize $H_{2}^{tt}$} 
  \put(-93,46){\footnotesize $H_{N}^{tt}$} 
  \put(-304,57){\footnotesize $F^{0}_{SR}$} 
  \put(-305,33){\footnotesize $F^{0}_{Rec}$} 
  \put(-259,70){\footnotesize $F^{1}_{SR}$} 
  \put(-256,32){\footnotesize $F^{1}_{Rec}$} 
  \put(-182,70){\footnotesize $F^{2}_{SR}$} 
  \put(-178,32){\footnotesize $F^{2}_{Rec}$} 
  \put(-106,70){\footnotesize $F^{N}_{SR}$} 
  \put(-103,32){\footnotesize $F^{N}_{Rec}$} 
  \put(-228,70){\footnotesize $F^{1}_{TT}$} 
  \put(-152,70){\footnotesize $F^{2}_{TT}$} 
  \put(-80,70){\footnotesize $F^{N}_{TT}$} 
  \caption{Overview of the proposed multi-task framework, including an SR branch, a reconstruction (Rec) branch, and a task transformer module.} 
  \label{fig1} 
\end{figure}

Our SR branch is used to enlarge the image from an undersampled and degenerated input $\hat{\mathbf{x}}_{LR}$. As shown in \figref{fig1}, for an input image of size $\frac{h}{s}\!\times\!\frac{w}{s}$ with artifacts, a convolutional layer is used to extract the shallow feature $F^{0}_{SR}$ of the SR branch. Then we send it to the backbone of EDSR~\cite{lim2017enhanced} to extract the SR features: 
$F^{1}_{SR}=H^{RB}_{SR_{1}}\left(F^{0}_{SR}\right)$, 
where $H^{RB}_{SR_{1}}$ represents the first Resblock in the SR branch. To enhance features from different tasks, we propose a task transformer module $H^{tt}$ (\S\ref{sec:tt}), which transfers the motion-artifacts-free representation to the SR branch. Formally, we have
 \begin{equation}
F^{i}_{TT}=H^{tt}_{{i}}\left(F^{i}_{SR}+F^{i}_{Rec}\right),  \quad i=1,2, \ldots, N,
\end{equation}
where $N$ is the number of $H^{tt}$, $F^{i}_{Rec}$ is the feature from the Rec branch (see Eq.~\eqref{eq:rec}), and $F^{i}_{SR}$ represents the $i$-th feature of the SR branch. The learned motion-artifacts-free representation $F^{i}_{TT}$ is then sent to the following Resblock:
\begin{equation}
F^{i+1}_{SR}=H^{RB}_{SR_{i+1}}\left(F^{i}_{TT}\right).
\end{equation}
Finally, a sub-pixel convolution $U_{\uparrow}$ is used as the upsampling module to generate the output ${\mathbf{x}}'$ of scale $h\!\times\!w$: ${\mathbf{x}}'=U_{\uparrow}\left(F^{N}_{SR}+F^{0}_{SR}\right)$. The whole branch is trained under the supervision of the  fully sampled HR image $\mathbf{x}$.

\subsubsection{Reconstruction Branch.}
As discussed above, only relying on the SR module is not sufficient for recovering a high-resolution and motion-corrected image when starting from an LR image with artifacts as input. Reconstruction, on the other hand, can restore a clear image with correct anatomical structure from an input with motion artifacts $\hat{\mathbf{x}}_{LR}$, because it is trained under the supervision of $\mathbf{x}_{LR}$. This means that reconstruction can effectively remove the artifacts introduced by the undersampled $k$-space, which is helpful for our final multi-task goal. By comparing the input and output of the Rec branch in \figref{fig1}, we can easily see that the Rec branch is more powerful in eliminating artifacts. For this branch, as shown in \figref{fig1}, we employ the same design as the SR branch to reduce the computational cost and generate high-quality results. We first use a convolutional layer to extract the shallow feature $F^{0}_{Rec}$ from the Rec branch. Then a series of $H^{RB}_{Rec_{i}}$ is used to extract the deep motion-corrected features
\begin{equation}\label{eq:rec}
F^{i}_{Rec}=H^{RB}_{Rec_{i}}\left(F^{i-1}_{Rec}\right), \quad i=1,2, \ldots, N,
\end{equation}
where $H^{RB}_{Rec_{i}}$ represents the $i$-th Resblocks, and $F^{i}_{Rec}$ represents the $i$-th feature of the Rec branch. The Rec branch is trained under the supervision of the LR motion-artifacts-free image ${\mathbf{x}}_{LR}$, aiming to remove the artifacts from the input. In our multi-task framework, the output of this branch is fused to the SR branch to obtain the final super-resolved, motion-artifact-free image.

\subsection{Task Transformer Module} \label{sec:tt}
\begin{figure}[!t]
\centering
  \includegraphics[width=0.9\textwidth]{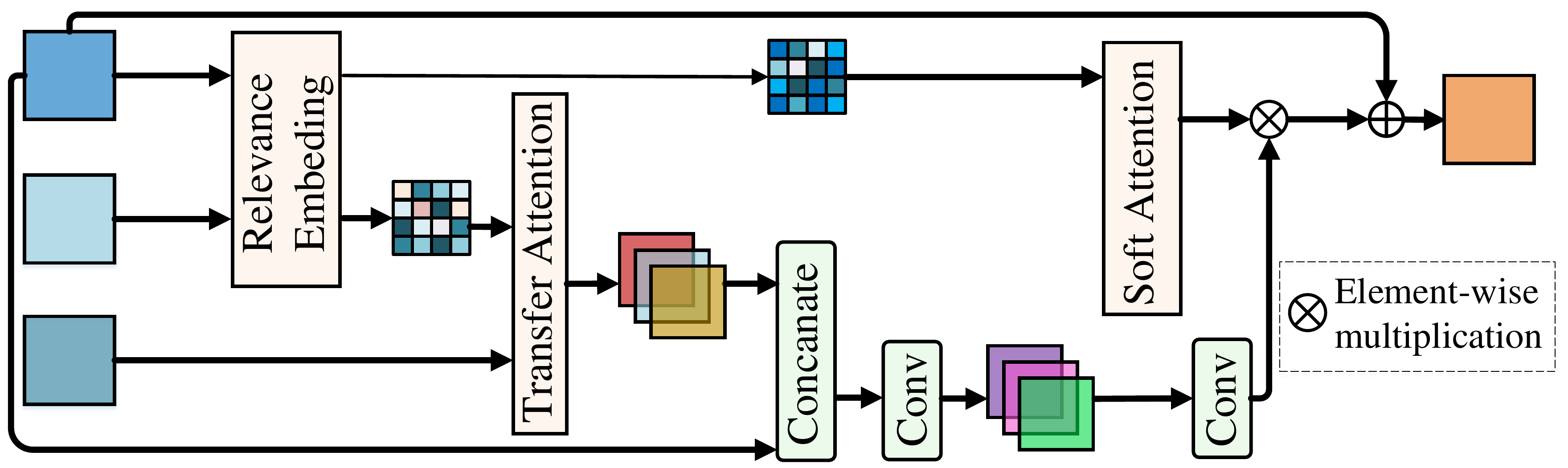}
  \put(-302,75){\footnotesize $Q$}
  \put(-303,46){\footnotesize $K$}
  \put(-302,18){\footnotesize $V$}
  \put(-289,68){\footnotesize $F^{i}_{SR}$}
  \put(-289,39){\footnotesize $F^{i}_{Rec}$}
  \put(-289,12){\footnotesize $F^{i}_{Rec} \uparrow \downarrow$}
  \put(-231,33){\footnotesize $T$}
  \put(-155,62){\footnotesize $S$}
  \put(-179,31){\footnotesize $C$}
  \put(-106,9){\footnotesize $Z$}
  \put(-24,67){\footnotesize $F_{TT}$}
  \caption{Architecture of the proposed task transformer module. $Q$ and $K$ are the features inherited from the SR and Rec branches, respectively. $V$ is the feature from the Rec branch with sequential upsampling and downsampling applied to it.} 
  \label{fig2} 
\end{figure}
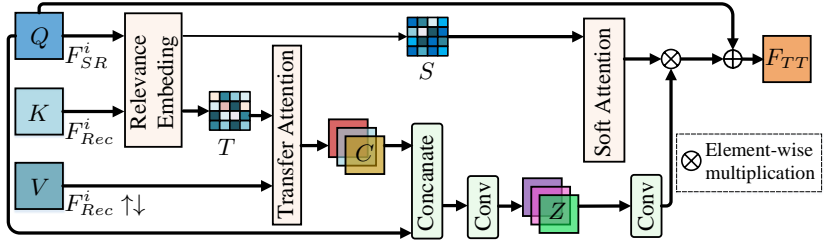

Since the Rec branch contains a stronger artifact removal capacity than the SR branch, we introduce a task transformer module to guide the SR branch to learn SR motion-artifacts-free representation from the Rec branch. Our task transformer module consists of three parts: a relevance embedding, a transfer attention for feature transfer and a soft attention for feature synthesis. As shown in \figref{fig2}, the features $F^{i}_{SR}$ and $F^{i}_{Rec}$ inherited from the SR and Rec branches are expressed as the query ($Q$) and key ($K$). The value ($V$) is the feature $F^{i}_{Rec}\uparrow\downarrow$ obtained by sequentially applying upsampling $\uparrow$ and downsampling $\downarrow$ on $F^{i}_{Rec}$ to make it domain-consistent with $Q$~\cite{yang2020learning}.

\subsubsection{Relevance Embedding.}
Relevance embedding aims to embed the relevance information from the Rec branch by estimating the similarity between $Q$ and $K$. To calculate the relevance $r_{i, j}$ between these two branches, we have
\begin{equation}
r_{i, j}=\left\langle\frac{q_{i}}{\left\|q_{i}\right\|}, \frac{k_{j}}{\left\|k_{j}\right\|}\right\rangle,
\end{equation}
where $q_{i}\left(i \in\left[1, \frac{h}{s}\!\times\!\frac{w}{s}\right]\right)$ and $k_{j}\left(j \in\left[1, \frac{h}{s}\!\times\!\frac{w}{s}\right]\right)$ are the patches of $Q$ and $K$, respectively.

\subsubsection{Transfer Attention.}
Our transfer attention module aims to transfer anatomical structure features from the Rec branch to the SR branch. Different from the traditional attention mechanism, we do not take a weighted sum of the reconstruction features for each query $q_i$, because this would result in blurred images for image restoration. To transfer features from the most relevant positions in the Rec branch for each $q_i$, we obtain a transfer attention map $T$ from the relevance $r_{i,j}$: $t_{i}={\arg \max_{j} } (r_{i, j})$, where $t_{i}\left(i \in\left[1, \frac{h}{2}\!\times\!\frac{w}{2}\right]\right)$ is the $i$-th element in $T$. We use the value of $t_i$ to represent the position in the Rec branch most relevant to the $i$-th position in the SR branch. 

To obtain the anatomical structure features $C$ without artifacts transferred from the Rec branch, an index selection operation is applied to the unfolded patches of $V$ using $t_i$ as the index: $c_{i}=v_{t_{i}}$, where $c_i$ represents the value of $C$ in the $i$-th position, which is equal to the $t_i$-th position of $V$.

\subsubsection{Soft Attention.}
To synthesize the features from the two branches in our model, we first concatenate $Q$ and $C$, and send them to a convolutional layer $Z=\operatorname{Conv}_{z}(\text {Concat}(C, Q))$. Then, we use a soft attention module to aggregate the synthetic features $Z$ and $Q$. To enhance the transferred anatomical structure information, we compute the soft attention map $S$ from $r_{i, j}$ to represent the confidence of the transferred structure features for each position in $C$: $s_{i}=\max_{j}(r_{i, j})$, where $s_{i}$ is the $i$-th position of the soft attention map $S$. To leverage more information from the SR branch, we first combine the synthetic feature $Z$ with the original feature of the SR branch $Q$. Then, the final output of the task transformer module is obtained as follows:
\begin{equation}
F_{\text {TT }}=Q \Moplus \operatorname{Conv}_{out}(Z) \Motimes S,
\end{equation}
where $\Moplus$ denotes the element-wise summation, $\Motimes$ denotes the element-wise multiplication, and $F_{\text {TT}}$ represents the final output of the task transformer module, which will be sent to the SR branch to restore a higher-quality, SR and motion-artifact-free image.

\section{Experiments}
\subsubsection{Datasets.}
We employ the public IXI dataset and a clinical brain MRI dataset to evaluate our method. The clinical dataset is scanned with fully sampling using a clinical 3T Siemens Magnetom Skyra system on 155 patients. The imaging protocol is as follows: matrix size 320$\times$320$\times$20, TR = 4511 ms, TE = 112.86 ms, field of view (FOV) = 230$\times$200 mm$^2$, turbo factor/echo train length TF = 16. For the IXI dataset, we exclude the first few slices of each volume since the frontal slices are much noisier than the others, making their distribution different. More details on the IXI dataset can be obtained from \url{http://brain-development.org/ixi-dataset/}. We split each dataset patient-wise into a ratio of 7:1:2 for training/validation/testing.

\subsubsection{Experimental Setup.}
For fair comparison, we implement four methods (two MRI reconstruction methods, ADMMNet~\cite{yang2016deep} and MICCAN~\cite{huang2019mri}, and two MRI SR methods, MGLRL~\cite{shi2018super} and Lyu \textit{et al.}~\cite{lyu2020mri}) with various sequential combinations, which we consider as baselines. These include: Com-A: ADMMNet-MGLRL, Com-B: ADMMNet-Lyu \textit{et al.}, Com-C: MICCAN-MGLRL, and Com-D: MICCAN-Lyu \textit{et al.}. The first model in each combination is used to remove artifacts, while the second is used to obtain higher-quality images. We implement our model in PyTorch using Adam with an initial learning rate of 5e-5, and train it on two NVIDIA Tesla V100 GPUs with 32GB of memory per card, for 50 epochs. Parameters $\alpha$ and $\beta$ are empirically set to 0.2 and 0.8, respectively. We use $N$ = 8 residual groups in our network. All the compared methods are retrained using their default parameter settings. 

\subsubsection{Experimental Results.}

We evaluate our multi-task model under 6$\times$ Cartesian acceleration with 2$\times$ and 4$\times$ enlargement, respectively. In Table~\ref{t1}, we report the average PSNR, SSIM and NMSE scores with respect to the baselines on the two datasets, where $\textit{w/o}$ $Rec$ and $\textit{w/o}$ $H^{tt}$ will discussed in the ablation study. On the IXI dataset, our T$^2$Net achieves a PSNR of up to 29.397 dB under 2$\times$ enlargement. Further, compared to the best sequential combination, we improve the PSNR from 21.895 to 28.659 dB under 4$\times$ enlargement. Moreover, with higher enlargement, the sequential combinations obtain worse scores, while our T$^2$Net still preserves excellent results. 
On the clinical dataset, our T$^2$Net again achieves significantly better results than all combinations, under both enlargement scales. This suggests that our model can effectively transfer anatomical structure features to the SR branch, and that this is beneficial to multi-task learning.

\begin{table*}[t]
 \centering
 \caption{Quantitative results on the two datasets under different enlargement scales. }
 \resizebox{\textwidth}{!}{
 \setlength\tabcolsep{1pt}
  \renewcommand\arraystretch{1.2}
 \begin{tabular}{r||ccc|ccc|ccc|ccc}
 \hline
  {Dataset} & \multicolumn{6}{c|}{IXI dataset}  &  \multicolumn{6}{c}{clinical dataset} \\ \cline{2-13}
  {Scale} & \multicolumn{3}{c|}{2$\times$} & \multicolumn{3}{c|}{4$\times$}  & \multicolumn{3}{c|}{2$\times$} & \multicolumn{3}{c}{4$\times$}  \\ \cline{2-13}
       & PSNR & SSIM & NMSE & PSNR & SSIM & NMSE & PSNR & SSIM & NMSE & PSNR & SSIM & NMSE  \\ \hline\hline
  Com-A &27.541 &0.801 &0.041 &21.111 &0.705 &0.178  &27.031 &0.764 &0.065 &26.169 &0.742 &0.079  \\
  Com-B &28.439 &0.847 &0.033 &21.323 &0.687 &0.170  &28.750 &0.816 &0.044 &27.539 &0.803 &0.058  \\
  Com-C &27.535 &0.802 &0.041 &21.696 &0.731 &0.156  &28.781 &0.765 &0.064 &26.197 &0.751 &0.079  \\
  Com-D &28.426 &0.847 &0.033 &21.895 &0.710 &0.149 &28.839 &0.817 &0.043 &27.700 &0.815 &0.056\\ \hline
  $\textit{w/o}$ $Rec$ &28.400 &0.809 &0.035  &25.952  &0.789 &0.091  &28.932   &0.802 &0.045 &28.601   &0.819 &0.044  \\
  $\textit{w/o}$ $H^{tt}$ &{28.700} &{0.856} & {0.031} &{26.692}  &{0.7730} &{0.089}  &{29.510} &{0.817} & {0.037} &{29.528}  &{0.821} & {0.037} \\ \hline
   T$^2$Net &\textbf{29.397} &\textbf{0.872} &\textbf{0.027} &\textbf{28.659}  &\textbf{0.850} & \textbf{0.032}  &\textbf{30.400}  &\textbf{0.841} &\textbf{0.030} &\textbf{30.252}  &\textbf{0.840} &\textbf{0.031} \\ \hline  
 \end{tabular}}
 \label{t1}
\end{table*}

We provide visual comparison results with corresponding error maps in Figure 3. The first two rows show the restored images and error maps from the IXI dataset with 6$\times$ Cartesian acceleration and 2$\times$ enlargement, while the last two rows are the results for the clinical dataset with 6$\times$ Cartesian acceleration and 4$\times$ enlargement. As we can see, the input has significant aliasing artifacts and loss of anatomical details. The sequential combination methods can improve the image quality, but are less effective than our multi-task methods. Our methods are obviously robust to aliasing artifacts and structural loss in the input. More importantly, at a high enlargement scale, our multi-task methods achieve much better results than the sequential combination methods. 

\begin{figure}[!t]
\centering
  \includegraphics[width=1\textwidth]{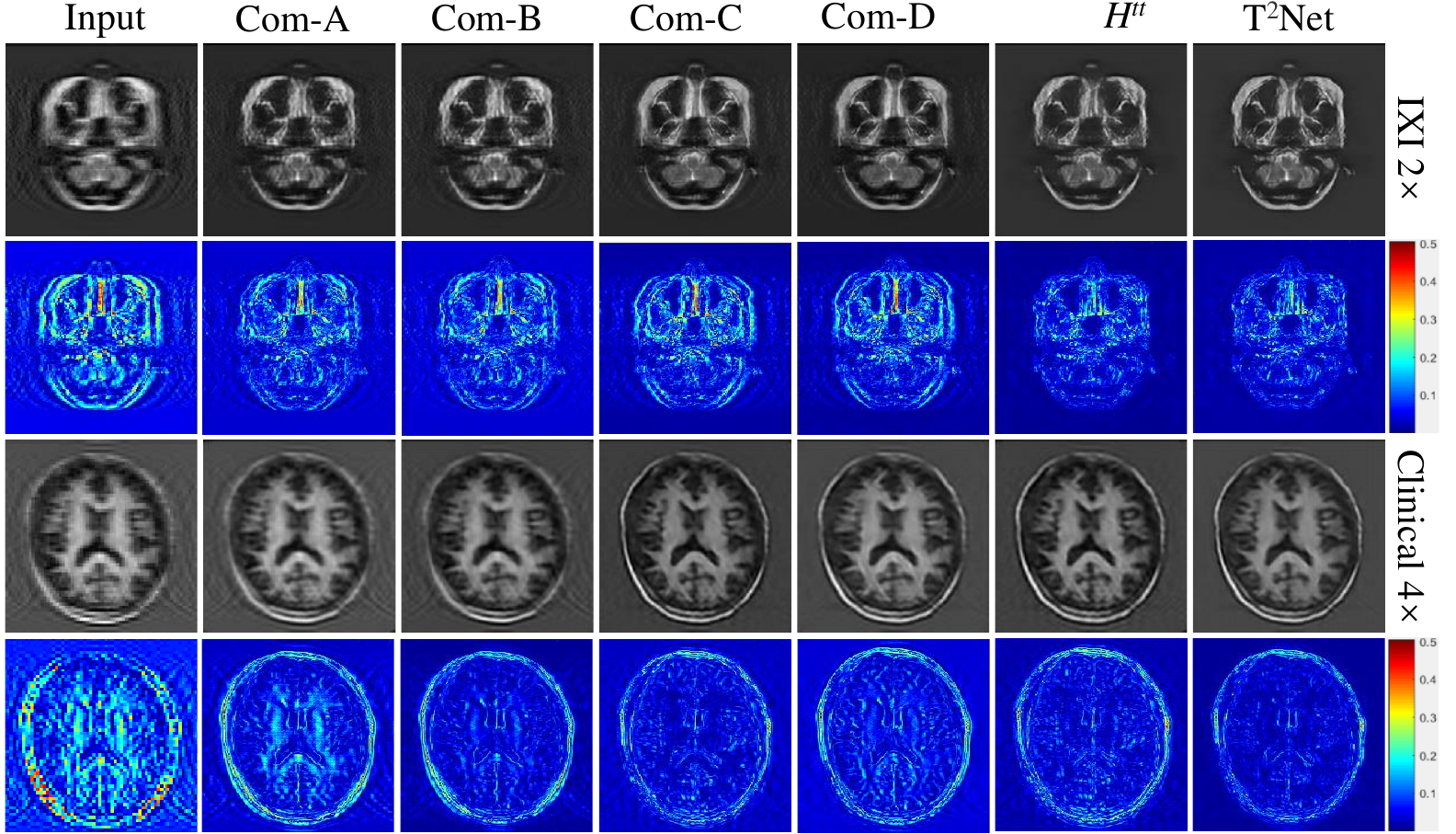} 
  \put(-103,192){\footnotesize $\textit{w/o}$}
  \caption{Visual comparison with error maps of different methods on the two datasets.} 
  \label{figure3} 
\end{figure}

  

\subsubsection{Ablation Study.}
We evaluate the effectiveness of the two branches and the task transformer module in our multi-task network. Without loss of generality, the restoration results of two key components, including the Rec branch and task transformer module, are evaluated under 6$\times$ acceleration and 2$\times$ as well as 4$\times$ enlargement. We summarize the component analysis in Table~\ref{t1}, where $\textit{w/o}$ $Rec$ indicates that only the SR branch is employed, while $\textit{w/o}$ $H^{tt}$ indicates that both branches are used but the task transformer module $H^{tt}$ is removed. As we can observe, $\textit{w/o}$ $Rec$ obtains the worst results, which indicates the importance of both branches in our method, as each contains task-specific features for the target image restoration. Moreover, we can also see that $\textit{w/o}$ $H^{tt}$ outperforms $\textit{w/o}$ $Rec$, demonstrating that transferring anatomical structure features to the target SR branch is necessary to achieve complementary representations. More importantly, our full T$^2$Net further improves the results on both datasets and all settings. This demonstrates the powerful capability of our $H^{tt}$ and two-branch structure in multi-task learning, which increases the model's flexibility to share complementary features for the restoration of higher-quality, super-resolved, and motion-artifacts-free images from highly undersampled and degenerated MRI data.

\section{Conclusion}
In this work, we focus on the multi-task learning of MRI reconstruction and super-resolution. For this purpose, we propose a novel end-to-end task transformer network (T$^2$Net) to transfer shared structure information to the task-specific branch for higher-quality and super-resolved reconstructions. 
Specifically, our model consists of two task-specific branches, \ie, a target branch for SR and auxiliary branch for reconstruction, together with a task transformer module to transfer anatomical structure information to the target branch. The proposed task transformer consists of a feature embedding, hard attention and soft attention to transfer and synthesize the final reconstructions with correct anatomical structure, whilst maintaining fine details and producing less blurring and artifacts. In the future, we will design a network to automatically learn the loss weights.

%
%
%
\bibliographystyle{splncs04}
\bibliography{bibliography}

\end{document}